\documentclass[%
 reprint,
 amsmath,amssymb,
 aps,
]{revtex4-2}

\usepackage{graphicx}
\usepackage{dcolumn}
\usepackage{bm}
\usepackage{subfigure}

\begin{document}

\preprint{APS/123-QED}

\title{Topological quantum phase transitions of anisotropic AFM Kitaev model driven by magnetic field}

\author{Shi-Qing Jia}
\affiliation{Key Laboratory of Materials Physics, Institute of Solid State Physics, HFIPS, Chinese Academy of
                          Sciences, Hefei 230031, China}
\affiliation{Science Island Branch of Graduate School, University of Science and Technology of China,
                                  Hefei 230026, China}

\author{Ya-Min Quan}
\email{Shi-Qing Jia and Ya-Min Quan contribute equally to this work}
\affiliation{Key Laboratory of Materials Physics, Institute of Solid State Physics, HFIPS, Chinese Academy of
                          Sciences, Hefei 230031, China}

\author{Liang-Jian Zou}
 \email{zou@theory.issp.ac.cn}
\affiliation{Key Laboratory of Materials Physics, Institute of Solid State Physics, HFIPS, Chinese Academy of
                           Sciences, Hefei 230031, China}
\affiliation{Science Island Branch of Graduate School, University of Science and Technology of China,
                                  Hefei 230026, China}

\author{Hai-Qing Lin}
 \email{haiqing0@csrc.ac.cn}
\affiliation{Beijing Computational Science Research Center, Beijing 100193, China 
}
\affiliation{Department of Physics, Beijing Normal University, Beijing 100875, China
}


\date{\today}

\begin{abstract}
We investigate the evolution of quantum spin liquid states of anisotropic Kitaev model
with the $[001]$ magnetic field by utilizing the finite-temperature Lanczos method (FTLM).
In this anisotropic antiferromagnetic Kitaev model with $K_{X}=K_{Y}$ and
$K_{X}+K_{Y}+K_{Z}=-3K$ ($K$ is the energy unit), due to the competition between anisotropy
and magnetic field, the  system emerges four exotic quantum phase transitions (QPTs) when $K_{Z}=-1.8K$
and $-1.4K$, while only two QPTs when $K_Z=-0.6K$.
In these magnetic-field tuning quantum phase transition points, the low-energy excitation
spectrums appear level crossover, and the specific heat, magnetic susceptibility and Wilson ratio
display anomalies; accordingly, the topological Chern number may also change.
These demonstrate that the anisotropic interacting Kitaev model with modulating
magnetic field displays more rich phase diagrams, in comparison with isotropic Kitaev model.

\begin{description}
\item[PACS numbers] 75.10.Kt, 75.10.Jm, 73.43.Nq, 71.15.Dx
\end{description}
\end{abstract}

\footnotetext[1]{Shi-Qing Jia and Ya-Min Quan contribute equally to this work}
\maketitle


\section{\label{sec:level1} Introduction} 

The search and study of the highly entangled quantum spin liquid (QSL) state have been the current frontier of condensed matter physics, especially after the Kitaev proposed an exactly solvable model on the two-dimension (2D) honeycomb lattice. The Kitaev model exhibits the gapless or gapped $Z_{2}$ QSL ground state related to the Majorana fermions resulting from spin fractionalization \cite{Kitaev2006}.  In this model the three nearest-neighbour (NN) bond-depending interactions are Ising-type anisotropic terms with the remarkable frustration and fluctuations, and the gapless or gapped QSL state appears when they are all equal or asymmetric\cite{Kitaev2006}.

So far, the candidate "Kitaev materials" are mainly implemented in the "spin-orbit entangled $j=1/2$ Mott insulators", which are $4d^{5}$ and $5d^{5}$ transition-metal oxide Mott insulators in the presence of strong relativistic spin-orbit coupling (SOC) and electronic correlations \cite{Pesin2010,Trebst2017}. The possible Kitaev materials are honeycomb iridium oxides $Na_{2}IrO_{3}$ \cite{Singh2010} and $Li_{2}IrO_{3}$ \cite{Kimchi2015} with $Ir^{4+}$ ($5d^{5}$) valence and Ru-based material $RuCl_{3}$ \cite{Plumb2014} with $Ru^{3+}$ ($4d^{5}$) valence. In these compounds, the exchange easy axis of main interactions depends on the spatial orientation of the exchange bond \cite{Khaliullin2005,Jackeli2009}, providing prototypes of the Kitaev couplings.

Actually, the Kitaev-type exchange interactions $K_{X}$, $K_{Y}$, and $K_{Z}$ are anisotropic in the NN $X$-, $Y$-, and $Z$-bonds in these realistic Kitaev materials. For example, the generalized spin Hamiltonian in $Na_{2}IrO_{3}$ is anisotropic, {\it i. e.}, $K_{X}=K_{Y}>K_{Z} \approx -30\,meV$. These parameters fitting from the ab-initio electronic structure calculations \cite{Yamaji2014,Suzuki2015,Yamaji2016} are consistent with those from the inelastic neutron scattering \cite{Choi2012,Banerjee2016} and the resonant X-ray magnetic scattering \cite{Liu2011} experiments well. Meanwhile, applying magnetic field may drive isotropic Kitaev model from a gapless QSL, through
a new $U(1)$ gapless QSL, to a polarized ferromagnetic (FM) phases \cite{Hickey2019}, or even an intermediate topological state with high Chern number\cite{Jiang2020}.
In the present anisotropic Kitaev interactions, external magnetic field brings about the competition with the anisotropic Kitaev coupling, may lead to more rich quantum phases comparing the isotropic Kiatev case.

However, similar to isotropic case, the anisotropic Kitaev model with the magnetic field is no longer exactly solvable.
Up to date several theoretical approaches, such as the thermal pure quantum (TPQ) state method \cite{Hickey2019}, the Majorana mean-field theory \cite{Liang2018,Nasu2018}, and the variational Monte Carlo calculations \cite{Liu2018,Jiang2020}, were proposed for the isotropic Kitaev model under external magnetic field.
To get deep insight into the anisotropic Kitaev model under the magnetic field modulation, following the ideas of the finite-temperature Lanczos method (FTLM) \cite{Komzsik2003,Prelovsek2013,Jaklic1994}, we develop the codes of FTLM  to obtain enough excited state information in finite temperature, and obtain the numerical results of the isotropic Kitaev model same to those by the TPQ method \cite{Hickey2019}.

In this paper, we perform the FTLM to gain an insight into the evolution of the QSL states in the anisotropic AFM Kitaev model with increasing magnetic field, especially focusing on critical points of the topological quantum phase transitions (QPT).
Based on the level crossover of low-energy excitation spectrums, the anomalies of the specific heat, magnetic susceptibility and Wilson ratio,
we find extra new QPTs accompanied with the change of Chern number, contrast with the isotropic Kitaev model,
suggesting that the anisotropic Kitaev coupling competing with magnetic field leads to new quantum phases
\cite{Po2017,Fulga2019,Molignini2021}. Combining the Majorana mean-field theory, the field-dependent topological Chern numbers are also presented.

The rest of the paper is organized as follows. In Sec. II, we define the anisotropic Kitaev model with an external magnetic field and outline the FTLM theory. In Sec. III and Sec.IV, we present the main results of the numerical calculations and discuss the essence and evolution of the QSL ground states. The conclusion of this paper is given in Sec. V.

\section{\label{sec:level2} Model Hamiltonian and Theoretical Methods}

\begin{figure}[htbp]
\centering
\subfigure[]{
\label{fig:state1a}
\includegraphics[angle=0, width=0.43 \columnwidth]{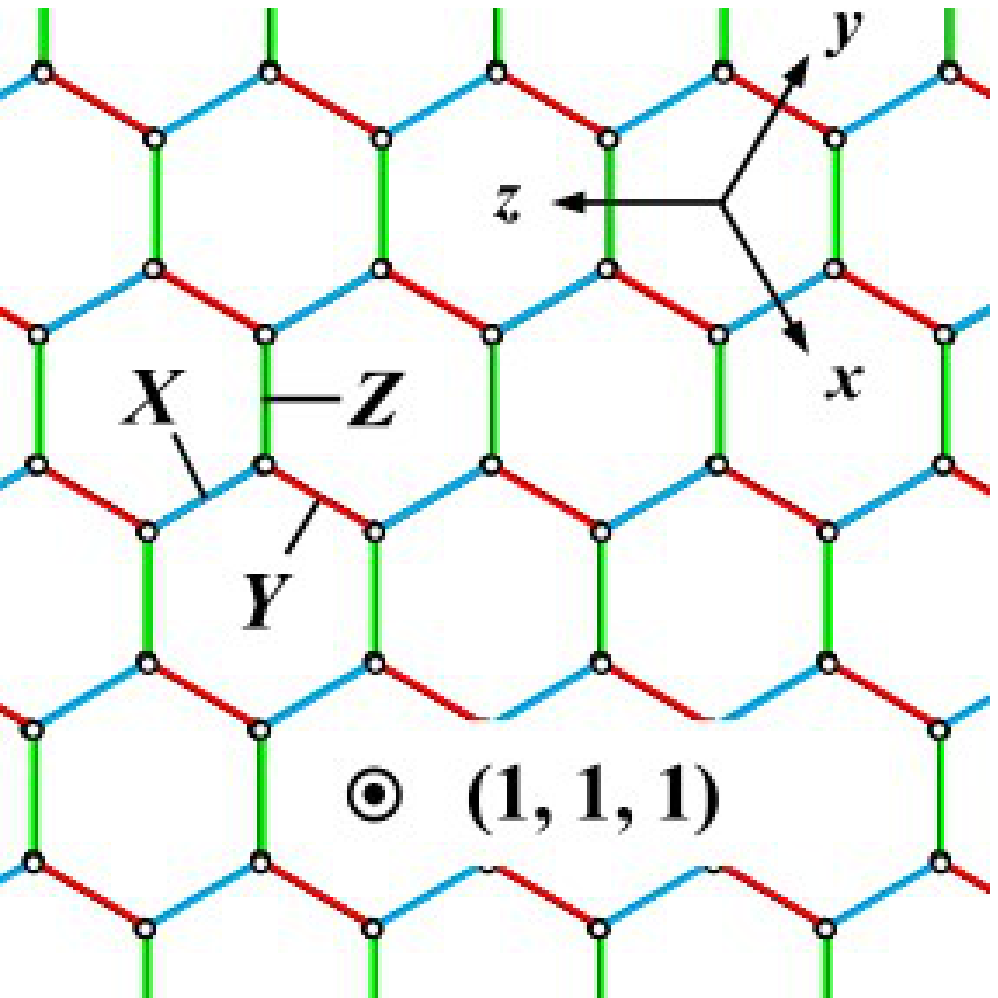}}
\hspace{0.25in}
\subfigure[]{
\label{fig:state1b}
\includegraphics[angle=0, width=0.43 \columnwidth]{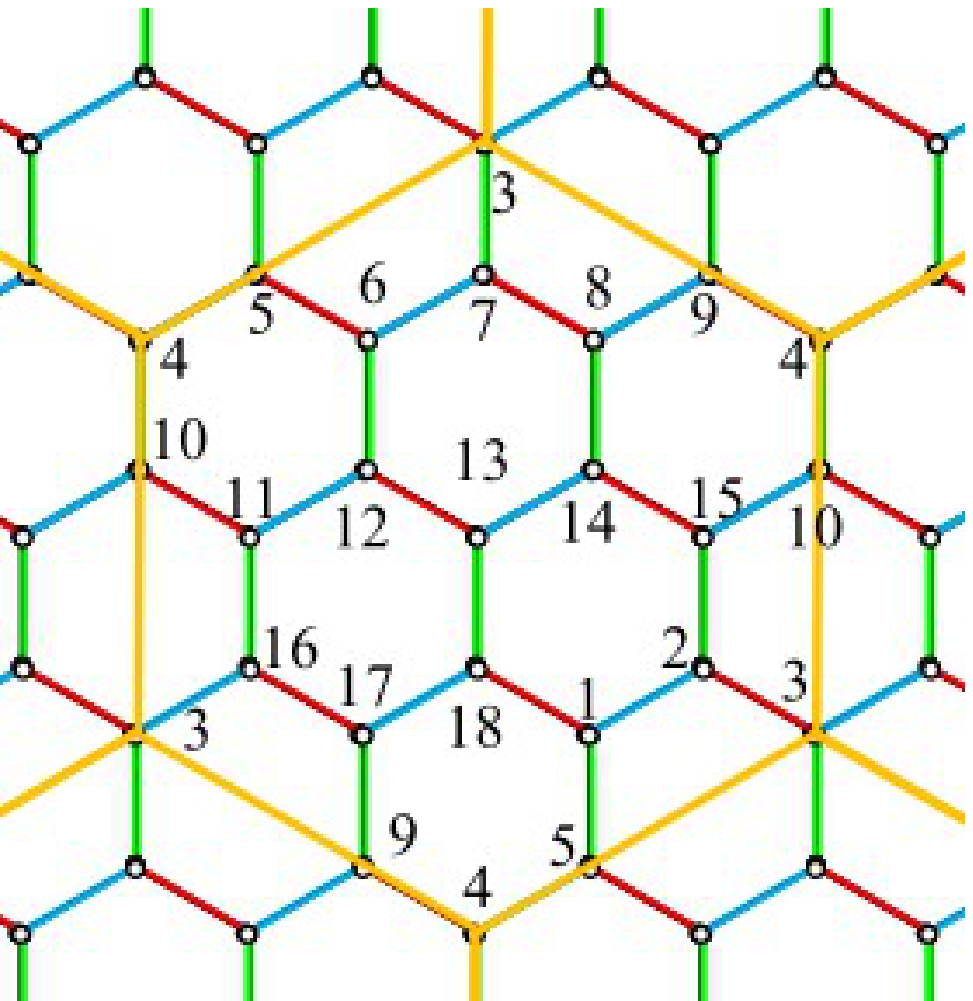}}
\caption{(Color online) (a) 2D honeycomb lattice structure located by Kitaev model. Blue, red, and green solid lines denote $X$, $Y$, and $Z$ bonds, respectively. $x$, $y$, and $z$ are the Kitaev axes of spins $S^{x}$, $S^{y}$, and $S^{z}$. The honeycomb plane is perpendicular to the $(x, y, z)=(1, 1, 1)$ direction. (b) The $N=18$-site cluster with $C_{6}$ rotational symmetry in the honeycomb lattice, we use it throughout this paper.}
\label{fig:state1}
\end{figure}

In this paper, the 2D honeycomb lattice of Kitaev model consists of two sublattices A and B, as illustrated in Fig.~\ref{fig:state1a}. Here we set that the Kitaev couplings in the NN $X$-, $Y$- and $Z$-bonds satisfy the conditions $K_{X}=K_{Y}$ and $K_{X}+K_{Y}+K_{Z}=-3K$ for the anisotropic Kitaev model. Throughout this paper the parameter $K$ is taken as the unity of energy, and the minus sign of -3K corresponds to the AFM coupling.

We start from the following Hamiltonian with the NN Kitaev couplings $K_{X}$, $K_{Y}$, $K_{Z}$ and the magnetic field $H_{z}$:
\begin{eqnarray}
\label{eq:Hamiltonian1}
 H&=&- K_{X}\sum_{\langle ij\rangle_{X}}S^{x}_{i} S^{x}_{j}
     - K_{Y}\sum_{\langle ij\rangle_{Y}}S^{y}_{i} S^{y}_{j}
     - K_{Z}\sum_{\langle ij\rangle_{Z}}S^{z}_{i} S^{z}_{j}   \nonumber\\
  & &{}-g\mu_{B}H_{z}\sum_{i}S^{z}_{i},
\end{eqnarray}
where the Land\'{e} factor $g=2$, $\mu_{B}$ is the Bohr magneton, and $H_{z}$ is external magnetic field in the $[001]$ direction of the spin frame. $\langle ij\rangle_{X}$, $\langle ij\rangle_{Y}$, and $\langle ij\rangle_{Z}$ limit the sum over the sites on the NN $X$, $Y$, and $Z$ directions, respectively. $S_{i}^{\alpha}$ ($\alpha=x, y, z$) represents the spin component at site $i$.

We investigate this Kitaev model by employing the FTLM theory \cite{Yamaji2016} on the $18$-site regular hexagon cluster with the periodic boundary condition and the full point group symmetry of the honeycomb lattice, such as the sixfold rotational symmetry ($C_{6}$), as illustrated in Fig.~\ref{fig:state1b}. FTLM algorithm is a quite effective method for finding the low energy eigenvalues of the Hamiltonian matrix with less arithmetic operations, high accuracy and large space size \cite{Komzsik2003,Prelovsek2013}. Firstly, based on the three-term recurrence relations, from a randomly selected basis vector we can derive a set of orthogonal basis vectors and construct a tridiagonal matrix in the Krylov subspace which is far smaller than the complete Hilbert space. Diagonalizing a series of similar tridiagonal matrices, we approach the low energy eigenstates of the Hamiltonian gradually, and further acquire the finite-temperature static expectation values, such as the total energy $E$, magnetic specific heat $C_{m}$, magnetic entropy $S_{m}$, magnetic susceptibility $\chi$ and Wilson ratio $R_{W}$ defined as follows.
\begin{eqnarray}
\label{eq:Hamiltonian2}
 E&=&\langle H\rangle,\quad  C_{m} = \frac{\langle H^{2}\rangle - \langle H\rangle^{2}}{t^{2}},\quad
      \chi = \frac{\langle \vec{S}^{2}\rangle - \langle \vec{S}\rangle^{2}}{t}   \nonumber\\
  & &R_{W} = \frac{4\pi^{2}}{3}\frac{\chi}{C_{m}/t},\quad S_{m} = \int_{0}^{t'}\frac{C_{m}}{t'}\,dt'
\end{eqnarray}
Throughout this paper we mainly present the numerical results for the $N=18$-site cluster, since it contains the regular hexagon boundary with full point-group symmetry without losing the generality.
In this study, we set the dimension of Krylov subspace to be $2^{11}$ for the $18$-site cluster with $2^{18}$ basis vectors. Throughout this paper, the dimensionless magnetic field is defined as $h_{z(x,y)}=g\mu_{B}H_{z(x,y)}/K$, and the dimensionless temperature is set as $t=k_{B}T/K$, with the energy unit $K=1.0$.

\section{\label{sec:level3} Numerical Results}

\begin{figure}[htbp]
\centering
\includegraphics[angle=0, width=0.85 \columnwidth]{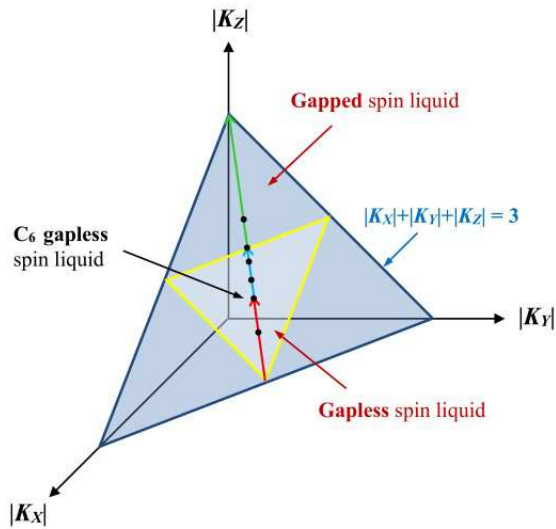}
\caption{(Color online) The variation range of Kitaev coupling strengths in the parametric phase diagram of the
 Kitaev model with the conditions $K_{X}=K_{Y}$ and $K_{X}+K_{Y}+K_{Z}=-3.0$ in the absence of magnetic field, marked by red, blue and green arrows.
 Six points marked with black dots, $K_{Z}=-1.8, -1.5, -1.4, -1.2, -1.0$, and $-0.6$, are chosen for our calculations.}
\label{fig:state2}
\end{figure}

In probing into the physical properties and evolution of the QSL states in the anisotropic AFM Kitaev model with the $[001]$ magnetic field, we take the range of the Kitaev couplings $K_Z$ along the line marked by red, blue, and green lines with arrows shown in Fig.~\ref{fig:state2}. We choose six representative points $K_{Z}=-1.8, -1.5, -1.4, -1.2, -1.0$ and $-0.6$, and mainly present the numerical data of three typical points with $K_{Z}=-1.8, -1.4$, and $-0.6$, respectively.

Along this line, in the absence of external magnetic field, the ground states of the Kitaev model experiences a gapped QSL for $|K_{Z}|>1.5$, and two different gapless QSLs for $1.0<|K_{Z}|\leq1.5$ and for $0.0<|K_{Z}|<1.0$, as shown in the green, blue and red segments of Fig.~\ref{fig:state2}, and the ground state at $|K_{Z}|=1.0$ is the gapless QSL with the $C_{6}$ rotational symmetry\cite{Kitaev2006}.

\subsection{\label{sec:level31} Low-energy excitation spectrums}

\begin{figure*}[htbp]
\centering
\includegraphics[angle=0, width=2.0\columnwidth]{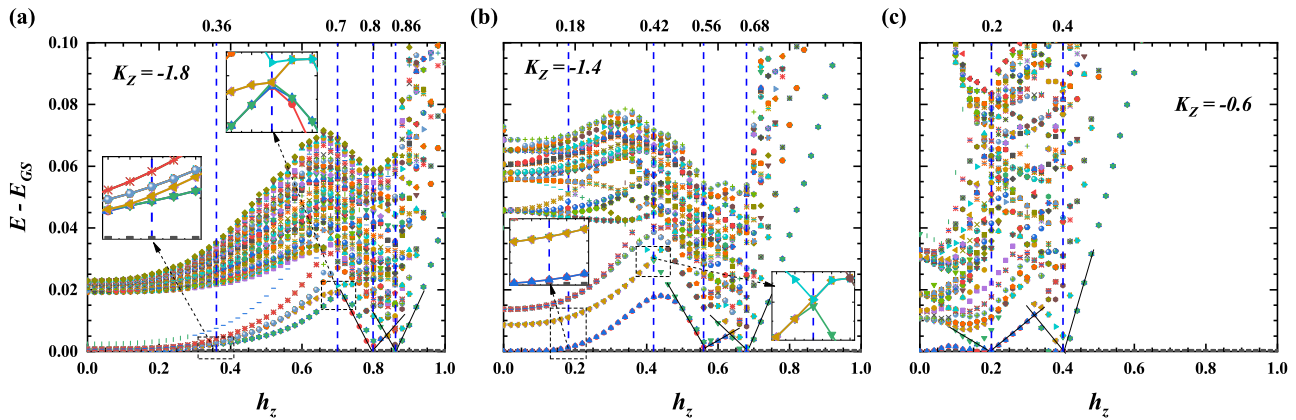}
\caption{(Color online) The low-energy excitation spectrums $(E-E_{GS})$ of anisotropic Kitaev model as functions of dimensionless magnetic field $h_{z}$ for different Kitaev couplings  $K_{Z}$=-1.8 (a), -1.4 (b), and $-0.6$ (c), respectively. The energy unit is K, N=18.}
\label{fig:state3}
\end{figure*}

Since the crossover of the low-energy excitation spectrums could identify the QPT points, we first present
the magnetic field dependences of low-energy excitation spectrums $(E-E_{GS})$ in the AFM anisotropic Kitaev model for different Kitaev couplings $K_{Z}=-1.8, -1.4$ and $-0.6$, respectively, as shown in Fig.~\ref{fig:state3}(a)-(c); here $E_{GS}$ is the energy eigenvalue of ground state.

At $|K_{Z}|=1.8$ in Fig.~\ref{fig:state3}(a), when the magnetic field increases from null, the energy degeneracy of the first and second excited states is removed at $h_{c1}$, as seen in the zoom image of Fig.~\ref{fig:state3}(a); with the further increasing field, a level crossover of the second and third excited states happens at $h_{c2}$, as also shown in the zoom image of Fig.~\ref{fig:state3}(a).
When $h_{z}>h_{c2}$, we observe another two level crossover points of the first excited state and ground state at $h_{c3}$ and $h_{c4}$, respectively.
At $h_{c3}$ and $h_{c4}$ one can see that the spin gaps close, similar to behavior of the isotropic Kitaev model under magnetic field\cite{Hickey2019}.
When $h_{z}>h_{c4}$, the energies of excited states increase nearly linearly with the open of a new spin gap.

When $|K_{Z}|=1.4$, similar to the $|K_{Z}|=1.8$ case, with the magnetic field increasing, the anisotropic Kitaev system also undergoes four QPTs points at the critical magnetic fields $h_{c1}$, $h_{c2}$, $h_{c3}$ and $h_{c4}$, respectively, as seen in Fig.~\ref{fig:state3}(b). Interestingly, in contrast to the system with $|K_{Z}|=1.8$, the phase boundaries $h_{c1}$-$h_{c4}$ shift to low fields with decreasing Kitaev coupling strength $|K_{Z}|$.
Moreover, when $|K_{Z}|=0.6$, only two level crossover points occur at $h_{c3}$ and $h_{c4}$, respectively, as shown in Fig.~\ref{fig:state3}(c).

\begin{figure}[htbp]
\centering
\includegraphics[angle=0, width=1.0 \columnwidth]{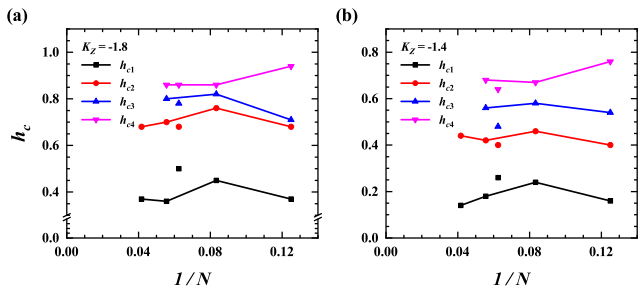}
\caption{(Color online) Finite-size scaling results of the critical magnetic fields $h_{c1}$, $h_{c2}$, $h_{c3}$, and $h_{c4}$ of the anisotropic AFM Kitaev model under the $[001]$ magnetic field for different Kitaev couplings $K_{Z}$=-1.8 (a), -1.4 (b), respectively, with the $8$-, $12$-, $16$-, $18$-, and $24$-site clusters.}
\label{fig:state4}
\end{figure}

One may question the stability of these QPT critical points when extending to infinite systems. Based on the low-energy spectrums, we display the finite-size extrapolation of the critical magnetic fields $h_{c1}$, $h_{c2}$, $h_{c3}$, and $h_{c4}$ of the anisotropic Kitaev model for different Kitaev couplings $K_{Z}=-1.8$ and $-1.4$, respectively, with the $8$-, $12$-, $16$-, $18$-, and $24$-site clusters. The numerical result is shown in Fig.~\ref{fig:state4}.
One can see that the four QPT critical points $h_{c1}$, $h_{c2}$, $h_{c3}$ and $h_{c4}$ approach finite values when the system-size $N$ becomes large enough, suggesting the robustness of these QPT critical points $h_{c1}-h_{c4}$ and the finite-size effect does not qualitatively change our conclusion. Thus, the numerical results we present here for the $18$-site cluster with the regular hexagon boundary and full point-group symmetry are qualitative reliability.

\subsection{\label{sec:level32} Magnetic specific heat}

\begin{figure*}[htbp]
\centering
\includegraphics[angle=0, width=2.05\columnwidth]{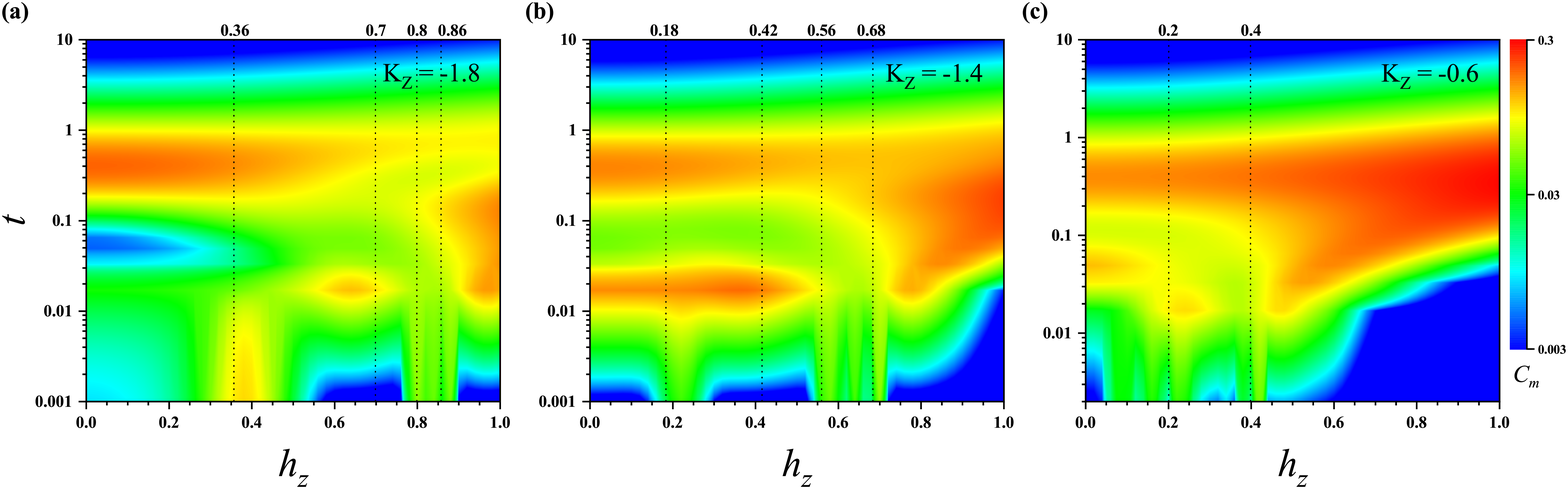}
\caption{(Color online) The phase diagrams of the anisotropic Kiatev systems based on the magnetic specific heats $C_{m}$ with the dimensionless magnetic field $h_{z}$ and temperature $t$ for different Kitaev couplings $K_{Z}$=-1.8 (a), -1.4 (b), and $-0.6$ (c), respectively.}
\label{fig:state5}
\end{figure*}

In order to investigate the evolution of the anisotropic Kitaev QSL state with magnetic field thoroughly, the temperature {\it vs} magnetic-field phase diagrams based on the magnetic specific heats $C_{m}$ for different Kitaev couplings $K_{Z}=-1.8, -1.4$ and $-0.6$ have been illustrated in Fig.~\ref{fig:state5}(a)-(c), respectively.

At $|K_{Z}|=1.8$ shown in Fig.~\ref{fig:state5}(a), we find that with the magnetic field increasing, four feature points are observed at low temperature, which correspond to the critical magnetic fields $h_{c1}- h_{c4}$ obtained in last subsection and in Fig.~\ref{fig:state3}(a).
One notices that at the critical magnetic fields $h_{c1}$, $h_{c3}$ and $h_{c4}$, the magnetic specific heats display three peaks when $h_z$ increases, respectively, demonstrating the features of the QPTs. These critical fields are robust with the temperature increasing when $t<0.01$. Whereas, at the critical magnetic fields $h_{c2}$, the magnetic specific heats display a dip, showing that $h_{c2}$ has feature different from another three critical fields.

Same to the critical fields of the spin excitation spectrums, when $|K_{Z}|=1.4$, with the magnetic field increasing, the anisotropic Kitaev system also goes through four QPT points at the critical magnetic fields $h_{c1}$, $h_{c2}$, $h_{c3}$ and $h_{c4}$, respectively, as seen in Fig.~\ref{fig:state5}(b). As expected, the phase boundaries $h_{c1}-h_{c4}$ move towards low fields with the decreasing Kitaev coupling strength $|K_{Z}|$. When $|K_{Z}|=0.6$, there are only two phase transition points at $h_{c3}$ and $h_{c4}$, respectively, as shown in Fig.~\ref{fig:state5}(c).

\begin{figure*}[htbp]
\centering
\includegraphics[angle=0, width=2.0 \columnwidth]{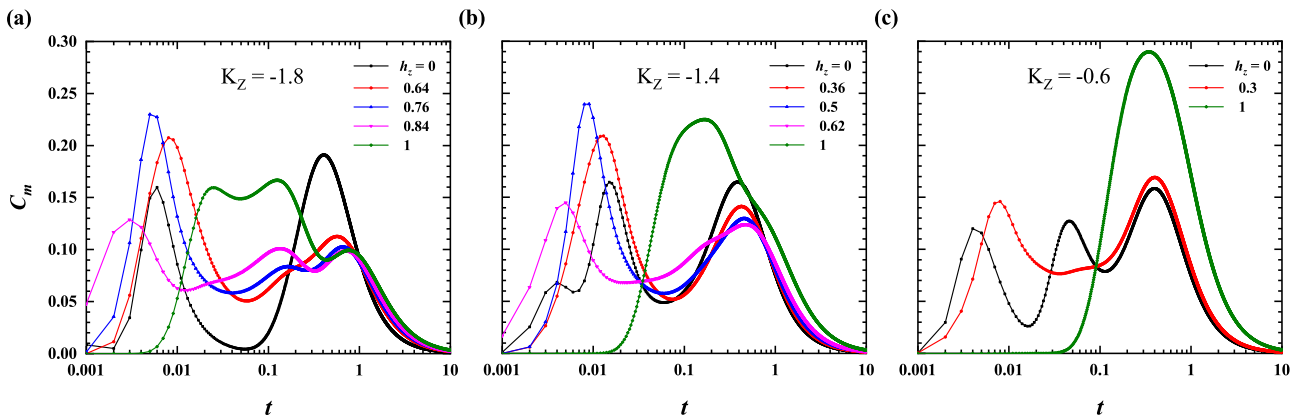}
\caption{(Color online) Temperature dependences of the magnetic specific heats $C_{m}$ with the dimensionless magnetic fields $h_{z}=0.0\thicksim1.0$ for $K_{Z}$=-1.8 (a), -1.4 (b), and -0.6 (c), respectively.}
\label{fig:state6}
\end{figure*}

Moreover, we also plot the magnetic specific heats as functions of the temperature for several typical quantum phases in the different regions separated by the critical magnetic fields, as shown in Fig.~\ref{fig:state6}(a)-(c).
When $|K_{Z}|=1.8$, as seen in Fig.~\ref{fig:state6}(a), we plot the $t$-dependent specific heat curves at five typical fields $h_{z}=0, 0.64, 0.76, 0.84$ and $1.0$, which could represent five different quantum phases among $0<h_{z}<1$.
At $h_{z}=0$, $C_m$ displays two peaks, {\it i.e.}, low-temperature (low-T) and high-temperature (high-T) peaks at the critical temperatures $t_{c1}$ and $t_{c2}$, associated with the local and itinerant Majorana fermion modes \cite{Yamaji2016}; and the spectral wight of the high-T peak is greater than that of the low-T peak;
when $h_{z}=0.64$, the spectral weight of the low-T peak in magnetic specific heat is significantly greater than the high-T one, which reveals a different quantum phase from the one at $h_{z}=0$;
then, at $h_{z}=0.76$, an intermediate peak appears at a feature temperature $t_{c3}$ between the low-T and high-T peaks, and it only exists in the present anisotropic Kitaev model, which may result from the interaction between the local and itinerant Majornana fermion modes in magnetic field; thus it implies a new quantum phase;
when $h_{z}=0.84$, in contrast to the case $h_{z}=0.76$, the spectral weights of the low-T and high-T peaks transfer to the intermediate peak, thus the intermediate peak raises; finally at $h_{z}=1.0$, the low-T peak merges into the intermediate peak, and the high-T peak tend to merge to them with the disappearance of the local and itinerant Majorana fermion modes.

When $|K_{Z}|=1.4$, as seen in Fig.~\ref{fig:state6}(b), we also plot $C_{m}-t$ curves at five typical fields $h_{z}=0, 0.36, 0.5, 0.62$ and $1.0$. The behavior of the magnetic specific heat of each quantum phase as the function of temperature is similar to the case of $|K_{Z}|=1.8$.
When $|K_{Z}|=0.6$, we choose three typical fields $h_{z}=0, 0.3$ and $1.0$ in the three regions separated by $h_{c3}$ and $h_{c4}$ to plot the $C_{m}-t$ curves, as seen in Fig.~\ref{fig:state6}(c). Accordingly, from
the magnetic specific heat behaviors we give rise to two QPT critical points at $h_{c3}$ and $h_{c4}$ and three quantum phases.

These results demonstrate that in the anisotropic Kitaev QSL, the thermodynamic behaviors of these quantum phases defined by the critical magnetic fields $h_{c3}$ and $h_{c4}$ for all $|K_{Z}|$, as well as $h_{c1}$ and $h_{c2}$ for $|K_{Z}|=1.4$ and $1.8$, are different, showing that these quantum phases are essentially different. As we show later, the four QPTs undergo  successively from the gapped QSL to gapless QSL, another gapless QSL, the $U(1)$ gapless QSL, and the saturated polarized FM phases for $|K_{Z}|=1.4$ and $1.8$; or from the gapless QSL to the $U(1)$ gapless QSL, and to the saturated polarized FM phases for $|K_{Z}|=0.6$, similar to the isotropic case.

\subsection{\label{sec:level33} Spin susceptibility}

\begin{figure*}[htbp]
\centering
\includegraphics[angle=0, width=2.05\columnwidth]{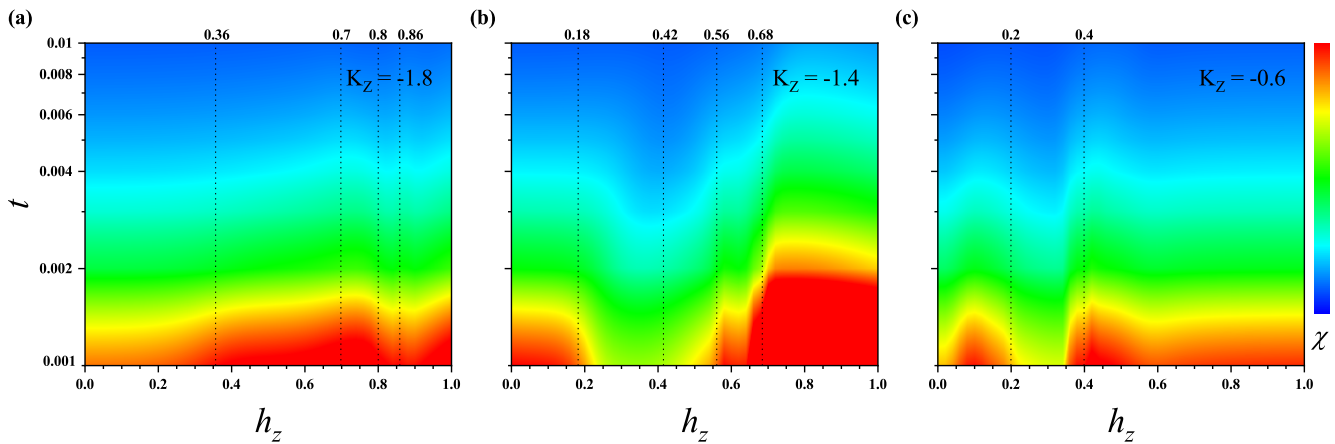}
\caption{(Color online)  The phase diagrams of the anisotropic Kitaev systems based on magnetic susceptibilities $\chi$ with the dimensionless magnetic field $h_{z}$ and temperature $t$ for different Kitaev couplings $K_{Z}=-1.8$ (a), $-1.4$ (b), and $-0.6$ (c), respectively.}
\label{fig:state7}
\end{figure*}

To further confirm the evolution of ground states of the anisotropic AFM Kitaev model, the temperature {\it vs} magnetic-field phase diagrams based on the magnetic susceptibilities $\chi$ for different Kitaev couplings $K_{Z}=-1.8, -1.4$ and $-0.6$ have been described in Fig.~\ref{fig:state7}(a)-(c).

When $|K_{Z}|=1.8$ and $t<0.01$, with the magnetic field increasing, the critical magnetic fields $h_{c1}$, $h_{c2}$, $h_{c3}$ and $h_{c4}$ indicated in the $t-h_z$ phase diagram of magnetic susceptibility shown in Fig.~\ref{fig:state7}(a) are not very clear as the discussed above.
In contrast, when $|K_{Z}|=1.4$, three QPT points at the critical magnetic fields $h_{c1}, h_{c3}, h_{c4}$ in the anisotropic Kitaev system are clear, as seen in Fig.~\ref{fig:state7}(b).
When $|K_{Z}|=0.6$, the two phase transition points at $h_{c3}$ and $h_{c4}$ are shown in Fig.~\ref{fig:state7}(c).

\begin{figure*}[htbp]
\centering
\includegraphics[angle=0, width=2.0 \columnwidth]{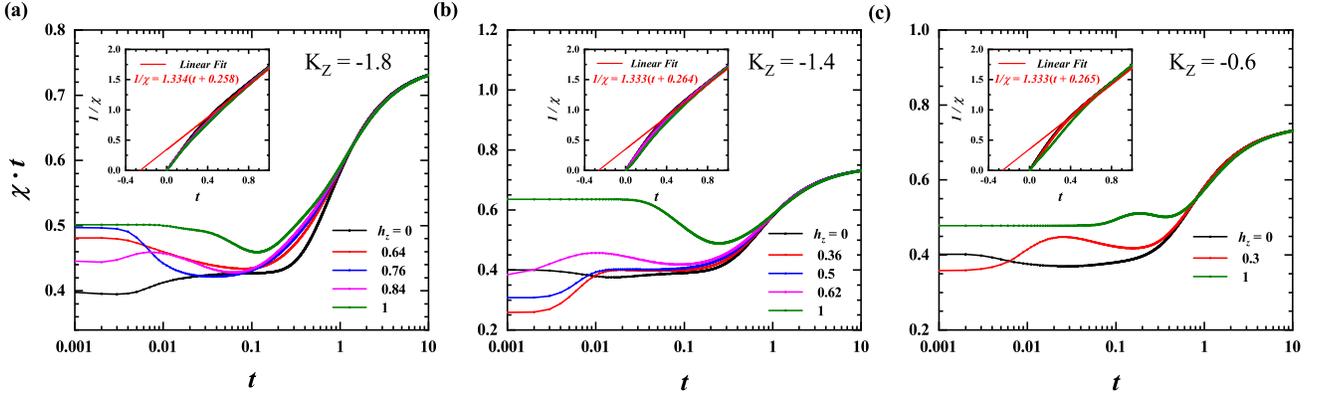}
\caption{(Color online) Temperature dependences of the products of the magnetic susceptibilities and temperature $\chi\cdot t$ and the inverse $1/\chi$ (Inset) at different magnetic fields for $K_{Z}$=-1.8 (a), -1.4 (b), and -0.6 (c), respectively.}
\label{fig:state8}
\end{figure*}

To further investigate the features of different quantum phases, we also plot the temperature dependences of the magnetic susceptibilities for the typical quantum phases as mentioned in the last section. Fig.~\ref{fig:state8} show the variations of magnetic susceptibility times temperature, $\chi\cdot t$, on increasing $t$; these curves display almost constants when $t<t_{c1}$ and converge to constants when $t\gg t_{c2}$.
From the Fig.~\ref{fig:state8}(a)-(c), we can see that these $\chi\cdot t$ curves of different quantum phases exhibit significantly different trends in the intermediate temperature range between $t_{c1}$ and $t_{c2}$, demonstrating the advantage of our FTLM for distinguishing the finite-temperature properties of different QSL phases.
In the temperature range $t_{c1}$ and $t_{c2}$, when $|K_{Z}|=1.8$, as seen in Fig.~\ref{fig:state8}(a), $\chi\cdot t$ show the kinks, or twists,  or dips, depending on $h_{z}=0, 0.64, 0.76$ and $0.84$ in different QSL phases;
when $h_{z}=1.0$, $\chi\cdot t$ exhibits a remarkable dip around $t_{c3}$, implying the presence of magnons in the polarized FM phase.
When $|K_{Z}|=1.4$ and $|K_{Z}|=0.6$, as shown in Fig.~\ref{fig:state8}(b) and (c), the $\chi\cdot t$ curves of the different quantum phases separated by $h_{c1-c4}$ are also distinguishable, similar to the counterparts at $|K_{Z}|=1.8$.

The inverses of the magnetic susceptibilities are also shown in the insets in Fig.~\ref{fig:state8}(a)-(c).
In the high-T region, all of the susceptibilities display a linear relationship $1/\chi=(T+\Theta_{W})/C_{A}$ approximately, where $\Theta_{W}$ is the Curie-Weiss temperature and $C_{A}$ is a constant, following the Curie-Weiss law of the AFM magnets very well. It is interested that all of the Curie-Weiss temperatures approximately equal to 0.25 for $K_{Z}=-1.8, -1.4$, and $-0.6$, seeming to stem from the constant condition $K_{X}+K_{Y}+K_{Z}=-3.0$.
In the low-T region with $t<0.3$, all the curves of inverses of magnetic susceptibilities bend anomalously and pass through the zero point, which violates the Curie-Weiss law and exhibits the features of the gapped or gapless QSL. On the whole, the field-driven QPT feature in the spin susceptibility is not so definite in comparison with that in magnetic specific heat.

\subsection{\label{sec:level34} Wilson ratio}

\begin{figure*}[htbp]
\centering
\includegraphics[angle=0, width=2.05\columnwidth]{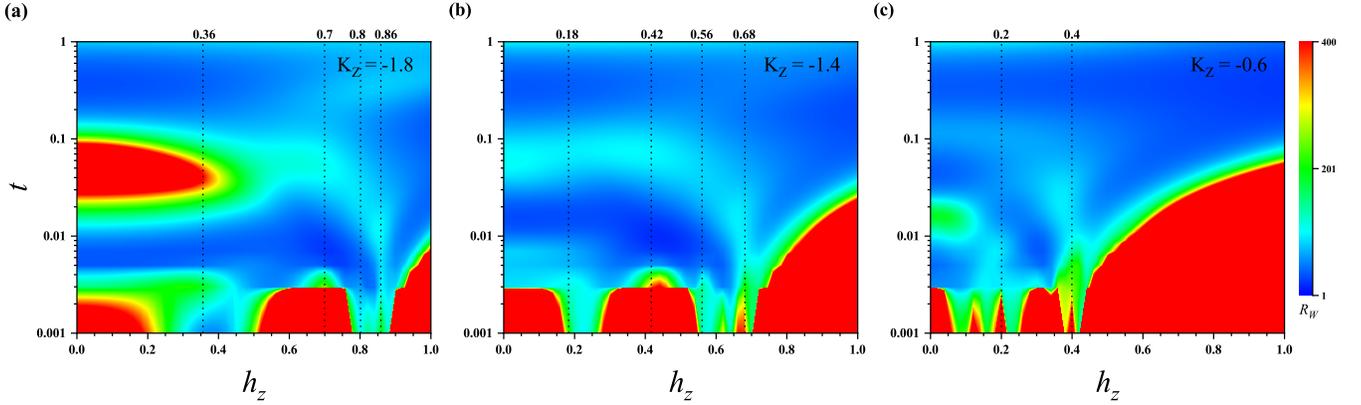}
\caption{(Color online)  The phase diagrams of the anisotropic Kitaev systems based on the Wilson ratio $R_{W}$ with the dimensionless magnetic field $h_{z}$ and temperature $t$ for different Kitaev couplings $K_{Z}$=-1.8 (a), -1.4 (b), and -0.6 (c), respectively.}
\label{fig:state9}
\end{figure*}

In order to further identify the QPT critical points of anisotropic AFM Kitaev model, we also plot the temperature {\it vs} magnetic-field phase diagrams based on the dimensionless Wilson ratios $R_{W}$. $R_{W}$ can represent the electronic correlation effects and quantify the spin correlations and fluctuations. For different Kitaev couplings $K_{Z}=-1.8, -1.4$ and $-0.6$, the phase diagrams are shown in Fig.~\ref{fig:state9}(a)-(c), respectively. From these phase diagrams we can see that all the Wilson ratios are larger than one, {\it i.e.} $R_{W} > 1$, indicating the strongly correlated features of the QSL ground states with the enhanced spin fluctuations.

When $|K_{Z}|=1.8$, with the increase of magnetic field, three remarkable feature points at $h_{c1}$, $h_{c3}$ and $h_{c4}$ are marked in Fig.~\ref{fig:state9}(a). These three critical magnetic fields lie in the low-$R_{W}$ regions since the low-T specific heats reach the maximum values. As a contrast, $h_{c2}$ falls in the high-$R_{W}$ region.
Similar situation also occurs when $|K_{Z}|=1.4$, with the increase of the magnetic field, the four QPT points at the critical magnetic fields $h_{c1}-h_{c4}$ are shown in Fig.~\ref{fig:state9}(b). When $|K_{Z}|=0.6$, the two phase transition points at $h_{c3}$ and $h_{c4}$ are displayed in Fig.~\ref{fig:state9}(c). Since the Wilson ratio is proportional to spin susceptibility over specific heat, similar to spin susceptibility, not all the QPT feature of the critical points are very clear in the phase diagrams based on the Wilson ratios.

\subsection{\label{sec:level35} Magnetic moments under the magnetic field}

\begin{figure*}[htbp]
\centering
\includegraphics[angle=0, width=2.0 \columnwidth]{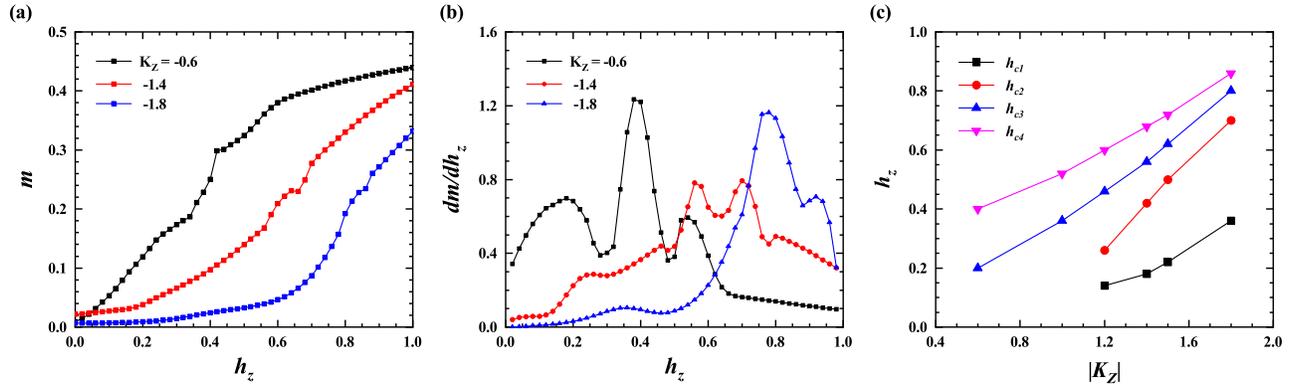} 
\caption{(Color online) Magnetic field $h_{z}$ dependences of the sublattice magnetic moments $m=\langle S^{z}\rangle$ (a) and its derivative  with respect to $h_{z}$, $dm/dh_{z}$ (b) at the zero temperature for different Kitaev couplings $K_{Z}=-0.6, -1.4$, and $-1.8$, respectively. (c) The critical magnetic fields $h_{c1}$, $h_{c2}$, $h_{c3}$, and $h_{c4}$ as functions of the Kitaev coupling strength $|K_{Z}|$.}
\label{fig:state10}
\end{figure*}

We also present the magnetic moments of sublattices $m=\langle S^{z}\rangle$ and their derivative respect to the magnetic field, $dm/dh_{z}$, as functions of magnetic field $h_{z}$ at zero temperature, which are illustrated in Fig.~\ref{fig:state10}(a) and (b). The magnetic moments have been rising up all the way with the increasing magnetic field because of magnetic polarization, among which we can find a few of turning points in these $m \thicksim h_{z}$ curves in Fig.~\ref{fig:state10}(a). Particularly it displays a small discontinuity at $h_{c4}$, seeming to be a first-order QPT.
More information could be found in the peaks of the zero-temperature susceptibility $dm/dh_{z} \thicksim h_{z}$ curves in Fig.~\ref{fig:state10}(b). The positions of these peaks almost one-to-one correspond to three distinct critical magnetic fields $h_{c1}$, $h_{c3}$ and $h_{c4}$ for $|K_{Z}|=1.8$ and $1.4$, and to two critical fields $h_{c3}$ and $h_{c4}$ for $|K_{Z}|=0.6$, respectively. Also, one finds a small shoulder for $|K_{Z}|=1.8$ and a small peak for $|K_{Z}|=1.4$ at $h_{c2}$, which may arise that it comes from the high-order level crossover in the low-energy excitations.

Note that the zero-temperature magnetic susceptibility $dm/dh_{z}$ should be quantitatively consistent with that obtained from the fluctuation-dissipation theorem in Eq.~\ref{eq:Hamiltonian2} in zero field limit in infinite system. Whereas, in the present finite-size clusters, due to the boundary effect, these two definitions display slight difference quantitatively in the positions of these QPT critical points.
When $h_{z}>h_{c4}$, we still observe an extra peak in $dm/dh_{z}$ for $|K_{Z}|=1.4$ or for $|K_{Z}|=0.6$. But in this case the system has already entered spin-polarized FM phase, this extra peak beyond $h_{c4}$ might come from the finite-size effect.

We also plot the dependences of the critical magnetic fields $h_{c1}$, $h_{c2}$, $h_{c3}$ and $h_{c4}$ on the Kitaev coupling strength $|K_{Z}|$, as shown in Fig.~\ref{fig:state10}(c). All these critical magnetic fields go up with the Kitaev coupling increasing. This arises from the fact that the total
spin gap increases with the lift of $|K_{Z}|$.

\section{\label{sec:level4}  Chern numbers and Discussions}

\begin{figure*}[htbp]
\centering
\includegraphics[angle=0, width=1.6\columnwidth]{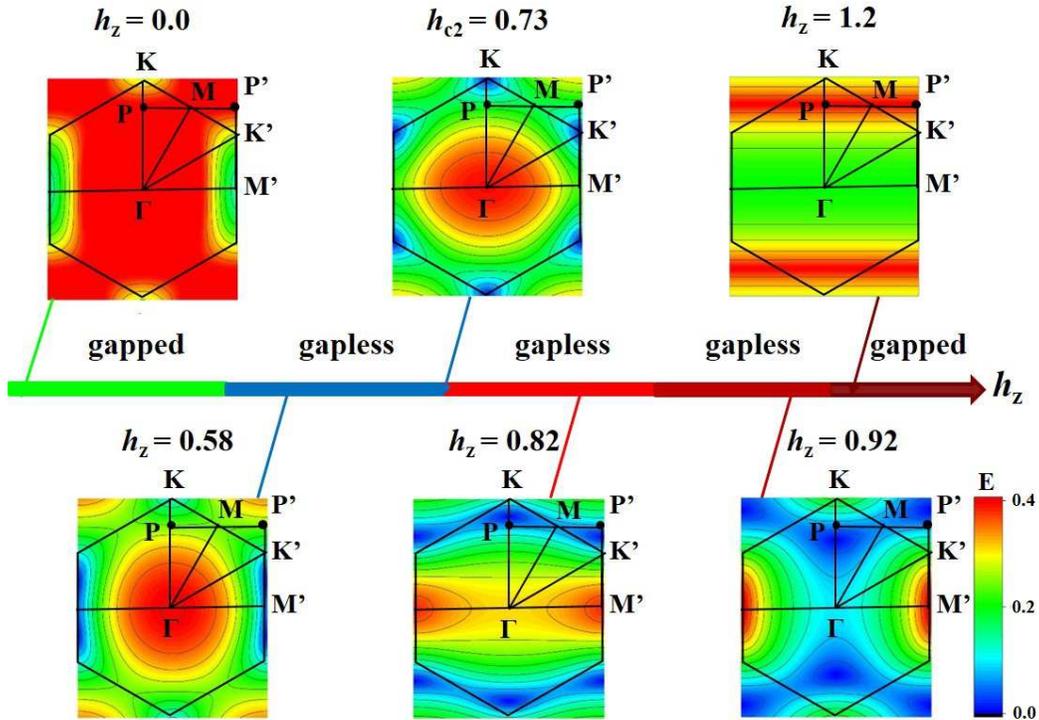}
\caption{(Color online)  The zero-temperature spinon spectrums of the anisotropic Kitaev systems as functions of the magnetic field $h_{z}$ for five typical phases at $h_{z}=0,0.58,0.82,0.92,1.2$, and $h_{c2}=0.73$, respectively. }
\label{fig:state11}
\end{figure*}

The competition of anisotropic Kitaev coupling and applied magnetic field resulting in at most five quantum phases could also seen in the Majorana fermion mean-field theory \cite{Liang2018,Nasu2018}. For example,  within the Majorana mean-field approximation, we find that for $K_{Z}=-1.8$, there are four QPT critical points at $h_{c1,c2,c3,c4}=0.51,0.73,0.87,0.96$, respectively, as shown in Fig.~\ref{fig:state11}. From which one finds that the spectral features of these five phases are distinctly different, further confirming the presence of four QPT critical points. The spinon energy dispersions in other parameter cases could be found in the {\it Supplementary Materials} \cite{Supple2021}

\begin{figure*}[htbp]
\centering
\includegraphics[angle=0, width=2.0\columnwidth]{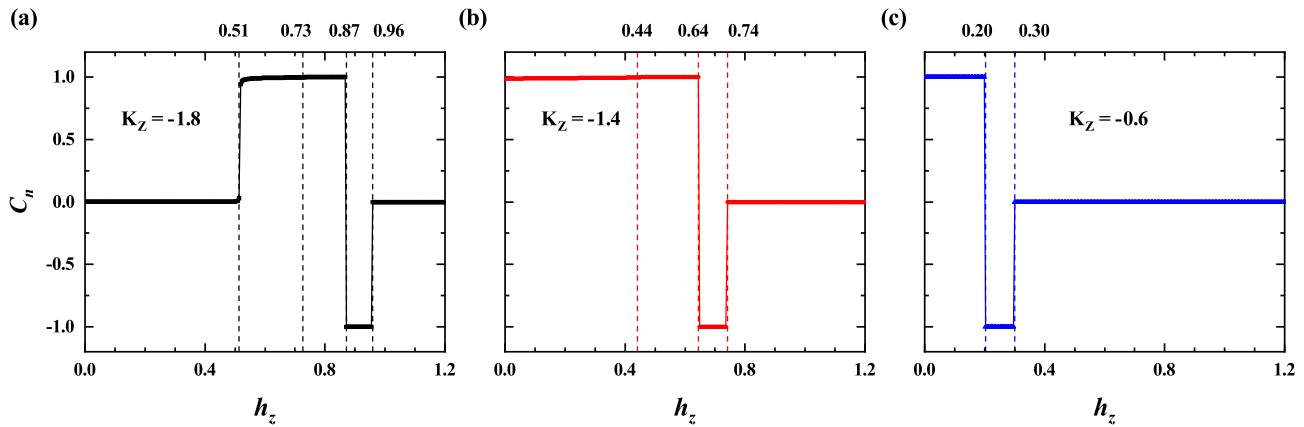}
\caption{(Color online)  The Chern numbers $C_{n}$ of the anisotropic Kitaev systems as functions of the magnetic field $h_{z}$ at $t=0$ for different Kitaev couplings $K_{Z}$=-1.8 (a), -1.4 (b), and -0.6 (c), respectively. Here the phase boundaries is determined through the spinon dispersions by the Majorana fermion mean-field method.}
\label{fig:state12}
\end{figure*}

Since the Kitaev model posses topological transitions, we further explore the evolution of the topological properties of the anisotropic Kitaev model with magnetic field, especially the topological QPTs.
With the help of the Majorana mean-field method, we could discuss the topological order with the continuous energy dispersions over the whole first Brillouin zone, and thus calculate the topological Chern numbers with the Wilson loop.
We display the magnetic field dependences of the Chern numbers $C_{n}$ at zero temperature, as shown in Fig.~\ref{fig:state12} (a)-(c).

When $|K_{Z}|=1.8$, from Fig.~\ref{fig:state12} (a), we can discover that the turning points of the Chern number correspond to the three critical magnetic fields one to one, which confirm the topological QPTs at $h_{c1}$, $h_{c3}$ and $h_{c4}$.
Specifically, at $h_{c1}$, the Chern number goes from 0 to 1 with the transition from the original gapped QSL to a gapless QSL; at $h_{c3}$, it changes from 1 to -1 with the transition from another gapless QSL to a new $U(1)$ gapless QSL \cite{Hickey2019}; at $h_{c4}$, it turns from -1 to 0 with the transition from the $U(1)$ gapless QSL to the polarized FM phase.
However, when $h_{z}$ passes through $h_{c2}$, the Chern number keeps 1 with the transition from the gapless QSL to another gapless QSL, and at $h_{c2}$ the system enters into a gapless QSL with six Weyl points at K and K', similar to the isotropic case.
These indicate that at $h_{c2}$ the topological order does not change, implying that $h_{c2}$ is not a topological QPT point, but a trivial one.

When $|K_{Z}|=1.4$, as seen in Fig.~\ref{fig:state12}(b), we observe the Chern number turns from 1, to -1, and finally to 0; the two topological QPT points occur at $h_{c3}$ and $h_{c4}$, respectively, and the QPTs at $h_{c1}$ and $h_{c2}$ are topological trivial.
From the spinon spectrums obtained by the mean-field method for $|K_{Z}|=1.4$, we find that through the critical magnetic fields $h_{c2}$, $h_{c3}$ and $h_{c4}$, the system transits from the original gapless QSL to another gapless QSL, to the $U(1)$ gapless QSL, and finally to the polarized FM phase.
And at $h_{c2}$, the system also goes into a gapless QSL with six Weyl points at K and K'.
Meanwhile, we do not observe the gap open again in the spinon spectrum at $h_{c1}$ for $|K_{Z}|=1.4$ in the Majorana mean-field approximation, different from the $|K_{Z}|=1.8$ case.
When $|K_{Z}|=0.6$, the two QPT points at $h_{c3}$ and $h_{c4}$ are also topological with the variations of the Chern number, from 1 to -1, and finally to 0, as seen in Fig.~\ref{fig:state12}(c). Accordingly, the system transits from the original gapless QSL to the new $U(1)$ gapless QSL, and to the polarized FM phase. Hence, most of the field-driven phase transitions in present anisotropic Kitaev model are topological QPTs.

From the preceding study one can see that different from the isotropic Kitaev model under magnetic field \cite{Liang2018,Nasu2018}, the anisotropic Kitaev coupling competing with applied magnetic field results in more rich quantum phases and QPTs.
Notice that the positions of the QPT critical points obtained from the level crossovers in the present FTLM approach are slightly different from those obtained from the Majorana mean-field approach, partially arising from the finite-size effect in the former, also partially from the underestimate of the spin fluctuations in the latter. Nevertheless, these critical fields lift up monotonically with the increasing Kitaev coupling $K_{Z}$, which originates from the fact that the energy gap of the system increases with the anisotropy.

\section{\label{sec:level5} Conclusion}

In summary, by employing the FTLM and combining the Majorana mean-field method, we study the nature and evolution of the QSL ground states in the anisotropic AFM Kitaev model with the $[001]$ magnetic field. In this Kitaev model with $K_{X}=K_{Y}$, $K_{X}+K_{Y}+K_{Z}=-3K$,  comparing with the isotropic Kitaev model, one finds that magnetic field may drive the appearance of new quantum phases, partially with the variation of the topological Chern number. Because of the competition between anisotropic term $K_{Z}$ and magnetic field $h_z$, the system exhibits three topological and one trivial QPTs when $K_{Z}=-1.8K$, and two topological and two trivial QPTs when $K_{Z}=-1.4K$; whereas the system with $K_Z=-0.6K$ displays only two topological QPTs, similar to the isotropic case.

Our results have shown that the anisotropic Kitaev coupling competing with magnetic field leads to intriguing rich phase diagram, and applied magnetic field could modulate and regulate the topologically different gapless and gapped QSL states, paving a way for the realization of quantum computations in realistic Kitaev materials \cite{Read1989,Kitaev2003,Kitaev2006}.
These results demonstrate that anisotropic Kitaev models may exhibit more interesting physics, and the nature of these new quantum QSL phases deserve further investigations.
\\

\begin{acknowledgments}
The author L. J. thanks the supports from the NSFC of China under Grant Nos.11774350 and 11474287, and H.Q. acknowledges financial support from NSAF U1930402 and NSFC 11734002. Numerical calculations were performed at the Center for Computational Science of CASHIPS and Tianhe II of CSRC.
\end{acknowledgments}

\providecommand{\noopsort}[1]{}\providecommand{\singleletter}[1]{#1}%
%



\end{document}